\documentstyle[prl,aps,multicol,psfig]{revtex}
\begin{document}
\input epsf
\draft

\title{ Information Capacity of a Hierarchical Neural Network }
\author{David Renato Carreta Dominguez}
\address{ Instituut voor Theoretische Fysica, K.U.Leuven\\
Celestijnenlaan, B-3001 Leuven, Belgium \\
e-mail: david@tfdec1.fys.kuleuven.ac.be}

\date{\today}
\maketitle

\begin{abstract}
The information conveyed by a hierarchical
attractor neural network is examined.
The network learns sets of correlated patterns (the examples) 
in the lowest level of the hierarchical tree and 
can categorize them at the upper levels.
A way to measure the non-extensive information content of the 
examples is formulated.
Curves showing the transition from a large retrieval information
to a large categorization information behavior,
when the number of examples increase,
are displayed.
The conditions for the maximal information are given
as functions of the correlation between examples and the
load of concepts.
Numerical simulations support the analytical results.
\end{abstract}


\begin{multicols}{2}
\narrowtext

\section{Introduction}

In the context of learning rules by perceptrons,
generalization by a neural network is the capability of 
correctly classify new patterns after some examples 
being taught to the network (see, e.g.,\cite{SS92}).
For attractor neural networks,
another type of generalization was suggested,
the categorization,
that emerges from an encoding stage where
a hierarchical tree of patterns is stored\cite{PV86}.
The ability of the network classifying the patterns 
on a lower level of the tree (i.e., the $examples$),
into categories defined by their ancestors (i.e., the $concepts$),
arises from the Hopfield model if the examples
are correlated with their concepts\cite{Fo90}.

A minimal number $S$ of examples for each concept is
necessary to start the categorization.
An extensive number of concepts is then learned by
memorizing finite sets of examples.
This was shown for networks of binary neurons with
fully-connected\cite{Mi91},\cite{KT93},
diluted\cite{CT95} or layered\cite{DT97} architectures,
and for analog\cite{ST92},
ternary\cite{DB97}, 
and non-monotonic\cite{Do96} neurons,
using Hebbian synapses.
Similar behavior was found for pseudo-inverse synapses\cite{NF98}.
Categorization is achieved through the appearance of
symmetric spurious states.
This ability to categorize start just when the capacity 
of the network recovering the original examples is lost,
because of the interference generated by 
their correlations.

As in most models for pattern recognition,
the adequate analysis of the memory capacity of this
networks require the tools of the information theory.
In the case of non-biased independent patterns,
one can avoid it
and measure the performance through the Hamming distance 
$D$ between the neuron and the retrieved pattern,
and the load capacity $\alpha$.
One scenario where $D$ and $\alpha$ are not enough to characterize
the system is that of sparse coded patterns\cite{DB98}.
Another one is that of dependent patterns.
This is case for categorization models, 
since the information conveyed by the examples is not extensive
in them.

Our goal in this work is to establish a reliable measure
for the capacity of retrieving examples and the categorization,
based in the information theory.
In the next section,
we define the model and its parameters.
After obtaining the expressions for the information capacity
in the section III,
in the section IV we study some special cases which present
the transition from a retrieval to a categorization phase.
Finally we conclude with some remarks in the section V.

\section{The Model}

Consider a network of $N$ binary neurons,
with states $\{\sigma_{i,t}\in\pm 1\}_{i=1}^{N}$ at time $t$.
The neurons states are updated in parallel
according to the deterministic rule

\begin{equation}
\sigma_{i,t+1}=\mbox{sign}(h_{i,t})\,\,;\,\,
h_{i,t}=\sum_{i(\neq j)}^{N}J_{ij}\sigma_{jt},
\label{2.sit}
\end{equation}
where $h_{i,t}$ is the local field of neuron $i$ at time $t$.
The elements of the Hebbian-like synaptic matrix 
between neurons $i$ and $j$ are given by

\begin{eqnarray}
 J_{ij}={1\over N}
\sum_{\mu}^{p}\sum_{\rho}^{S}
\eta^{\mu\rho}_{i}\eta^{\mu\rho}_{j},
\label{2.Jij}
\end{eqnarray}
where $\{\eta^{\mu\rho}_{i}\}_{\rho=1}^{S}$
are the $examples$ of the $concept$ $\xi^{\mu}_{i}$.
The concepts are independent identically distributed 
random variables ($IIDRV$),
$\{\xi^{\mu}_{i}=\pm 1\}_{i=1}^{p}$,
with equal probability.

In the encoding stage,
the examples are built from the concepts,
according to the stochastic process:

\begin{equation}
p(\eta^{\mu\rho}_{i}|\xi^{\mu}_{i})=
b\delta(\eta^{\mu\rho}_{i}-\xi^{\mu}_{i}) +
(1-b) \delta(|\eta^{\mu\rho}_{i}|^{2}-1),
\label{2.pem}
\end{equation}
where $b= \langle\eta^{\mu\rho}_{i} \xi^{\mu}_{i}\rangle $
gives the correlation between the ancestors (the concepts)
and the descendants (the examples) of this tree of patterns.
The second delta of this conditional distribution
gives the component of the examples which is
independent on the concepts.
This process can equivalently be formulated as
$\eta^{\mu\rho}_{i}=\xi^{\mu}_{i}\lambda^{\mu\rho}_{i}$,
where the $biased$ IIDRV $\lambda^{\mu\rho}_{i}$
are distributed according to

\begin{equation}
p_{B}(\lambda^{\mu\rho}_{i}) = 
B_{+}\delta(\lambda^{\mu\rho}_{i}-1) +
B_{-}\delta(\lambda^{\mu\rho}_{i}+1),
\label{2.pBl}
\end{equation}
with $B_{\pm}=(1\pm b)/2$.

The macroscopic parameters which describe the state of the
network are
the $retrieval$ and
the $categorization$ $overlaps$,
respectively:

\begin{eqnarray}
m^{\mu\rho}_{Nt} \equiv {1\over N} 
\sum_{j} \eta^{\mu\rho}_{j}\sigma_{jt} ,\,\,
M^{\mu}_{Nt} \equiv 
{1\over N} \sum_{j}\xi^{\mu}_{j}\sigma_{jt}.
\label{2.mmr}
\end{eqnarray}
In the thermodynamic limit, 
the qualities of the retrieval and of the categorization 
can be measured by taking the $\lim N\to\infty$ of the 
overlaps for a single concept, say $\mu=1$,
which give

\begin{eqnarray}
m^{1\rho}_{t}\equiv
\langle\eta^{1\rho}\mbox{sign}[h_{t-1}]\rangle,\,\,
M^{1}_{t}\equiv
\langle\xi^{1}\mbox{sign}[h_{t-1}]\rangle,
\label{2.m1r}
\end{eqnarray}
where the brackets mean averages over the set of examples
$\eta^{1\rho}$ and the local field $h_{t-1}$ for a single neuron.

The $generalization$ $error$\cite{SS92},\cite{Fo90}
can be defined as 
$E^{1}_{t} \equiv 
\langle|\sigma_{t}-\xi^{1}|^{2}\rangle
= 1-M^{1}_{t}$,
as a function of the categorization overlap.
The stationary states are given by macroscopic
overlaps with examples of a given concept,
say $m^{1\rho}_{\infty}\equiv m^{1\rho}$,
and microscopic remaining overlaps $\nu>1$,
$m^{\nu\rho}\sim 1/\sqrt{N}$.
The general solution is represented by a retrieval overlap
with a single example, 
say $m^{11}\equiv m$,
and the $quasi-symmetric$ overlaps with the other examples,
$m^{1\rho}\equiv m^{S},\,\rho>1$.
In the retrieval phase one have 
$m\sim 1$, $m^{S}\sim b^{2}$, 
while in the categorization phase the stable state is
$m=m^{S}\sim b$, 
which may leads to a large categorization overlap,
$M^{1}_{\infty}\equiv M\sim 1$.
In the following we will consider a situation where the network
have relaxed to the equilibrium states,
so we can drop the time $t$ on the parameters.

\section{Information Capacities}

In this section we describe a way to measure the storage of
information by the network in the retrieving and categorizing 
regimes.
There are two types of information to be extracted from the
patterns in these network:
the retrieval information and the categorization information.
The former is that which can be conveyed from the examples to
the neurons,
while the latter is that which can be conveyed from the concepts.
In each case one must calculate the information entropy of the
pattern distributions,
$ H[\{\xi^{\mu}_{i}\}_{i,\mu}^{N,p}] =
-\sum_{\{\xi^{\mu}_{i}\}}
p(\{\xi^{\mu}_{i}\})
\log[p(\{\xi^{\mu}_{i}\})] $,
and
$ H[\{\eta^{\mu\rho}_{i}\}_{i,\mu,\rho}^{N,p,S}] =
-\sum_{\{\eta^{\mu\rho}_{i}\}}
p(\{\eta^{\mu\rho}_{i}\})
\log[p(\{\eta^{\mu\rho}_{i}\})] $,
where 
$p(\{\xi^{\mu}_{i}\})$ and 
$p(\{\eta^{\mu\rho}_{i}\})$
are the concepts and examples joint probability distributions,
respectively.

The categorization information can be easily measured by computing
the categorization overlap of a single concept, $M$,
and its entropy.
Since the concepts 
$\{\xi^{\mu}_{i}\}_{i,\mu}^{N,p}$ are IIDRV,
their probability distribution is factorial,
$p(\{\xi^{\mu}_{i}\}_{i,\mu}^{N,p}) =
\prod_{i,\mu}^{N,p} p(\xi^{\mu}_{i})$.
Thus,
the entropy of the concepts is extensive,
$H[\{\xi^{\mu}_{i}\}_{i,\mu}^{N,p}] =
\sum_{\mu,i}^{p,N} H[\xi^{\mu}_{i}] = pN H[\xi]$,
where the entropy of a single concept on a single neuron is
$H[\xi]=\log(2)$.
As we study binary patterns, 
we shall use base-2 logarithm in order to count information
in bits,
then we have $H[\xi]=1$.
The equivocation in the categorization can be evaluated by the
square of the overlap, 
in such a way that no information is transmitted by the concepts
if $M=0$ and the information is maximal if $M=\pm 1$,
reminding that the information is symmetric in this overlap,
because an inverted concept $\sigma_{i}=-\xi_{i}$ carries 
the same information than $\sigma_{i}=\xi_{i}$.
Therefore,
the total categorization information is
$I_{C}=  pN M^{2} H[\xi]$,
and the categorization information (per synapse) is

\begin{equation}
i_{C}= \alpha M^{2} .
\label{3.iCa}
\end{equation}

The retrieval information can be similarly measured,
by computing the retrieval overlap and the entropy of the examples,
since this entropy can also be factorized as
$p(\{\eta^{\mu\rho}_{i}\}_{i,\mu,\rho}^{N,p,S}) =
\prod_{i,\mu}^{N,p}
p(\{\eta^{\mu\rho}_{i}\}_{\rho}^{S})$,
such that the entropy is extensive in the concepts and
in the neurons,
$ H[\{\eta^{\mu\rho}_{i}\}_{i,\mu,\rho}^{N,p,S}] =
\sum_{\mu,i}^{p,N} 
H[\{\eta^{\mu\rho}_{i}\}_{\rho}^{S}] 
=pN H[\{\eta^{\rho}\}_{\rho}^{S}]$.
Thus it is enough to calculate the entropy of a 
set of examples of a single concept,
$\{\eta^{\rho}\}_{\rho}^{S} \equiv 
\{\eta^{\mu\rho}_{i}\}_{\rho}^{S}$,
on a single neuron,
to get the entropy of the whole set 
$\{\eta^{\mu\rho}_{i}\}_{i,\mu,\rho}^{N,p,S}$.

On the other hand,
$\{\eta^{\rho}\}_{\rho}^{S}$
is $not$ a set of IIDRV, so
$p(\{\eta^{\rho}\}_{\rho}^{S})$
is $not$ factorizable in example probabilities,
and the entropy is $not$ extensive in the examples,
$ H[\{\eta^{\rho}\}_{\rho}^{S}] \neq
\sum_{\rho}^{S} H[\eta^{\rho}] $.
So the retrieval information is not the naive one,  
$i_{R}\neq\alpha S$.

Let $\{\eta^{\rho}\} \equiv \{\eta^{\rho}\}_{\rho}^{S}$ 
be a set of examples of a given concept on a given neuron.
In calculating $p(\{\eta^{\rho}\})$ we proceed as it follows:
we take the conditional probability of the examples given
the concept,
$p(\{\eta^{\rho}\}|\xi)$,
from Eq.(\ref{2.pem}),
and average it on the distribution of $\xi$,

\begin{equation}
p(\{\eta^{\rho}\}) = \langle 
p(\{\eta^{\rho}\}|\xi) 
\rangle_{\xi} =
\prod_{\rho=1}^{S}  
{p_{B}(\eta^{\rho}) + p_{B}(-\eta^{\rho}) \over 2} ,
\label{3.per}
\end{equation}
where $p_{B}$ is the probability distribution in
Eq.(\ref{2.pBl}).
After expanding this product,
we calculated the entropy of this distribution,
obtaining

\begin{eqnarray}
&&H[\{\eta^{\rho}\}]=
-\sum^{S}_{k=0} C^{S}_{k} A_{k} \log (A_{k});\nonumber\\
&&A_{k}= [B_{+}^{k}B_{-}^{S-k}+B_{-}^{k}B_{+}^{S-k}]/2,
\label{3.Her}
\end{eqnarray}
where $C^{S}_{k}$ are the combinatorial numbers.

In evaluating the equivocation in the retrieval, 
we here have to multiply this entropy by the square of the
retrieval overlap of a single example.
Since we have to subtract the information due to the categorization,
and the overlaps between examples and their concepts are 
$b= \langle\eta^{\rho} \xi\rangle $,
we estimate the total retrieval information as
$I_{R}=  pN (m-bM)^{2} H[\{\eta^{\rho}\}]$.
Therefore the retrieval information (per synapse) is

\begin{equation}
i_{R}= \alpha (m-bM)^{2} H[\{\eta^{\rho}\}] .
\label{3.iRa}
\end{equation}

Although other measures for the informations could be used,
they must be monotonous functions of those we consider in the 
Eqs.(\ref{3.iCa}),(\ref{3.iRa}).
These have, nevertheless, 
the advantage that both are equivalently scaled and
they can be directly compared to each other.

\section{Results}

We present now the equilibrium states for the networks which 
are used to obtain the retrieval and categorization informations.
This states are studied for two systems:
an asymptotic network ($N\to\infty$), 
for which analytical stationary equations were derived\cite{Fo90}
and finite-sized systems,
for which simulations of the dynamics in Eq.(\ref{2.sit})
are carried on.
While the information measures obtained in the previous section 
are functions of asymptotic parameters,
$M$ and $m$,
the results from simulation use the overlaps in Eqs.(\ref{2.mmr}).

\subsection{Asymptotic network}

First we study the stationary states of the overlaps in 
Eqs.(\ref{2.mmr}),
in the thermodynamic limit $N\to\infty$.
Using the Hebbian synapses in Eq.(\ref{2.Jij})
in the dynamics in Eq.(\ref{2.sit}),
taking the local field at the fixed point, 
and averaging over the distribution of a single example,
one get:

\begin{eqnarray}
m=&& \sum_{k=0}^{S-1} p_{S}(k)
\int_{-\infty}^{\infty} Dz
[ B_{+}G_{+} - B_{-}G_{-} ] , \nonumber\\
M=&& \sum_{k=0}^{S-1} p_{S}(k)
\int_{-\infty}^{\infty} Dz
[ B_{+}G_{+} + B_{-}G_{-} ] , \nonumber\\
m^{S}=&& \sum_{k=0}^{S-1} p_{S}(k) 
{x_{S}\over S-1} \int_{-\infty}^{\infty} Dz
[ B_{+} G_{+} + B_{-} G_{-} ] ,
\label{4.msk}
\end{eqnarray}
for the retrieval, categorization and quasi-symmetric overlap, 
respectively.
Here

\begin{equation}
G_{\pm}= \mbox{sign}[x_{S}m^{S}\pm m+z\sqrt{\alpha r}],
\label{4.Gpm}
\end{equation}
with
$x_{S}\equiv\sum_{\rho=2}^{S}\lambda_{\rho}\equiv 2k-(S-1)$.
and the averages are over the remaining 
$S-1$ examples from the first concept,
and the remaining $p-1$ concepts.
The first is the binomial variable
$x_{S}=2k-(S-1)$,
distributed according to

\begin{equation}
p_{S}(k)=
C^{S-1}_{k} B_{+}^{k}B_{-}^{S-1-k},
\label{4.pSk}
\end{equation}
the last is a Gaussian noise,
distributed according to

\begin{equation}
Dz = { dz\over\sqrt{2\pi} }
e^{ -{z^{2}\over 2} }.
\label{4.Dod}
\end{equation}
In the present case of a fully-connected network,
there is a strong feedback in the dynamics, 
but an expression for the variance of the noise
can be obtained using a replica symmetric approach\cite{Fo90},\cite{KT93},

\begin{equation}
r=s{ [1-C(1-b^{2})(1-b^{2}+sb^{2})]^{2} +(s-1)b^{4}
\over [1-C(1-b^{2})]^{2} [1-C(1-b^{2}+sb^{2})]^{2} },
\label{4.rs1}
\end{equation}
with

\begin{equation}
C = {1\over\sqrt{\alpha r}} 
\sum_{k=0}^{S-1} p_{S}(k) 
\int_{-\infty}^{\infty} Dz\,\, z
[B_{+} G_{+} + B_{-} G_{-} ] .
\label{4.C1a}
\end{equation}

We have to solve this Eqs.(\ref{4.msk})-(\ref{4.C1a}),
then we introduce the overlaps in the expressions for the
informations, Eqs.(\ref{3.iCa}),(\ref{3.iRa}).
These analytical results for the information 
are then presented in comparison 
with the results from simulation.

\subsection{Simulation}

The simulations we have performed are for networks of 
$N=5,000$ and $N=10^4$ neurons,
which are updated in parallel according to the dynamics in 
Eq.(\ref{2.sit}),
up to $t=10$ time steps,
or when the overlaps converge.
Thus we have $almost$ stationary states in most cases,
except when a state of non-information is obtained,
for which the times of convergence are typically much larger.

The capacity is analized as a function of the two parameters 
of loading of the network:
the rate of loading of concepts, $\alpha=p/N$,
and the number of examples per concept, $S$.
The sample averages are taken over an interval in 
$\ln(S)$ or in $\alpha$.
When simulating the information as a function of $S$, 
we generate first the concepts and then store consecutively the
examples of each concept.
When simulating the information as a function of $\alpha$, 
we generate the $S$ examples of the concept generated at each
step of the learning.

The network is trained then storing examples,
while the retrieval and categorization overlaps are monotorized.
For a fixed $\alpha$,
it is expected that increasing $S$ the network pass from a
regime where the retrieving information is large to another 
where the categorizing information increase up to saturation
in a upper bound.
This behavior is seen in Fig.1,
where the overlaps, 
as well as the informations,
are plotted as a function of $\ln(S)$, 
with a correlation $b=0.3$,
for a loading of concepts $\alpha=0.01$.
When more and more examples are learned,
the retrieval information increases until a maximum at $S_{R}=7$,
then it falls down.
After a while when no information is transmitted,
the network reach, at $S_{C}\sim 33$,
the categorization phase,
where the categorization information jumps to a higher value.
It continues to increase untill it saturates in $i_{C}=0.01$,
when the network reach $M\sim 1$ after $S\sim 90$.
The retrieval information capacity of the network is
$i_{R}\sim 0.06$.
The asymptotic theory for $N\to\infty$ fits quite well the
simulation for $N=10^4$, 
except in the region of no information.
This is due to the finite number of steps used in the dynamical
simulation, $t=10$,
while the convergence to the fixed-point there is very slow.

A case with a larger load of concepts, 
$\alpha=0.04$, is plotted in Fig.2.
Although now the network can only retrieve well the examples 
up to $S=3$, it has $i_{R}\sim 0.10$. 
Then there is a large waiting period where the informations stay
close to zero,
up to $S_{C}\sim 74$,
when the categorization information jumps to $i_{C}\sim 0.04$,
which is much larger than in the case $\alpha=0.01$.

Comparing this with a network with larger correlation,
$b=0.4$, plotted in Fig.3,
we observe that the network can store with a larger overlap
only $S_{R}=2$ examples,
with a maximal retrieval information $i_{R}\sim 0.05$, 
which is somewhat smaller than the naive $S\alpha\sim 0.08$.
However the categorization information  
approaches its saturation value $i_{C}\sim 0.04$ much faster, 
only $S\sim 30$ examples must be learned.
We have checked that for larger load of concepts
($\alpha\geq 0.06$) the 
categorization information is larger than the retrieval information.
Also we verified the for higher correlations ($b\geq 0.6$) the 
categorization information can be the larger one,
even for small load $\alpha\sim 0.01$,
while for smaller correlations ($b\leq 0.2$) 
the retrieval information is always the larger one.

For a fixed $S$,
one expects that increasing $\alpha$ 
the categorization information (if $b$ or $S$ are large enough)
increases up to a maximum value after which it decreases until
it becomes zero at a critical $\alpha$.
This behavior can be seen in Fig.4,
where the case $b=0.2$, $S=170$ is plotted.
We verified that the larger the values of $b$,
the higher are the maxima of $i_{C}$, 
and less examples are need. 
We also observed that the retrieval information have a 
similar non-monotonic behavior if $b$ or $S$ are small.

\section{Conclusion}

The information conveyed by the categorization model was studied.
It was shown that the transition from the retrieval phase to the
categorization phase carries together a transition in the
information:
the retrieval information decreases when the network is 
oversaturated with examples,
and after a period of resting,
the categorization information increases.

It is interesting to note that,
although neither the retrieval nor the categorization informations
surpasses the usual Hopfield model, 
($S=1$, $b=1$),
which is $i_{R}\sim 0.13$ at $\alpha=0.135$,
the fact that the network can return to behave as an associative
memory after a long period of $resting$ between $S_{R}<S<S_{C}$
is an advantage with respect to Hopfield network.
It is also of worth of note that the retrieval information can
still be relatively large, as we see in Fig.2,
a quotation which was not observed before in any work about
the categorization model in the literature.

The simulation results fit very well with the theoretical
in both retrieval and categorization regimes,
showing that almost no effect of finite-size is present,
but the time of convergence in the resting period must be
much larger than that used in this work.

Both expressions for the information of the retrieval and
of the categorization in Eqs.(\ref{3.iRa}-\ref{3.iCa})
are not claimed to be exact.
They are approximations for a more precise measure,
the $mutual$ $information$\cite{Sh48} between neuron and patterns,
$ {\cal I}[\sigma,\xi] = H[\xi] - 
\langle H[\sigma|\xi] \rangle_{\xi} $,
where $H[\sigma|\xi]$ is the conditional entropy.
Since we know that the conditional probability of the neuron,
given the concept state, is 
$ p(\sigma|\xi)= (1+M\sigma\xi)\delta(|\sigma|^{2}-1) $,
we can replace the categorization information by

\begin{equation}
{\cal I}[\sigma,\xi]=
{1+M\over 2}\ln(1+M)+
{1-M\over 2}\ln(1-M).
\label{5.I1+}
\end{equation}
This quantity gives the degree of information the
neuron can catch from the concept.
However we prefer to use the estimation in
Eq.(\ref{3.iCa}) to compare with the retrieval information
with the same precision.

Finally we hope that the present approach to the information
content of a neural network of correlated patterns can be
used in the context of more general architectures and 
learning rules.
A more general distribution of the 
$\lambda^{\mu\rho}_{i}$\cite{DT96}
may also deserves some attention.

{\bf Acknowledgments}

This work was financially supported by 
the Research Fund of the K.U.Leuven (grant OT/94/9).\\


\vspace{-0cm}
\begin{figure}[t]
\begin{center}
\epsfysize=10cm
\leavevmode
\epsfbox[1 1 600 500]{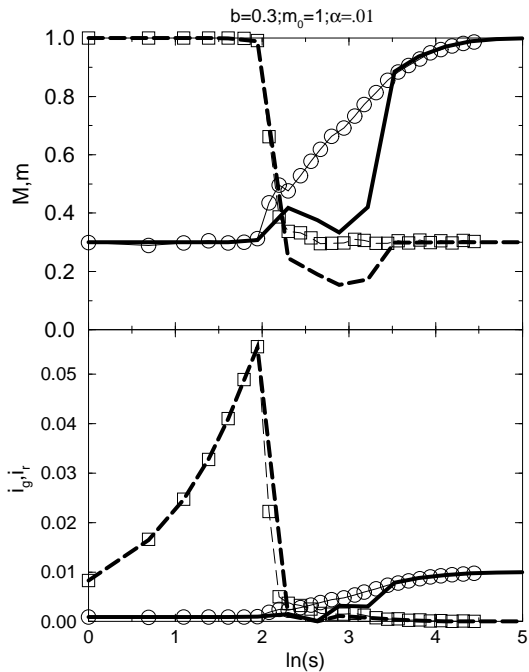}
\end{center}
\caption{ The overlaps (Top) and Informations (Bottom)
as functions of $\ln(S)$, for $b=0.3$ and $\alpha=0.01$.
The squares (circles) are the simulation results for 
retrieval (categorization) for $N=10^4$, $t=10$;
the dashed (solid) curves are the asymptotic results. }
\label{1}
\end{figure}

\vspace{-0cm}
\begin{figure}[t]
\begin{center}
\epsfysize=10cm
\leavevmode
\epsfbox[1 1 600 500]{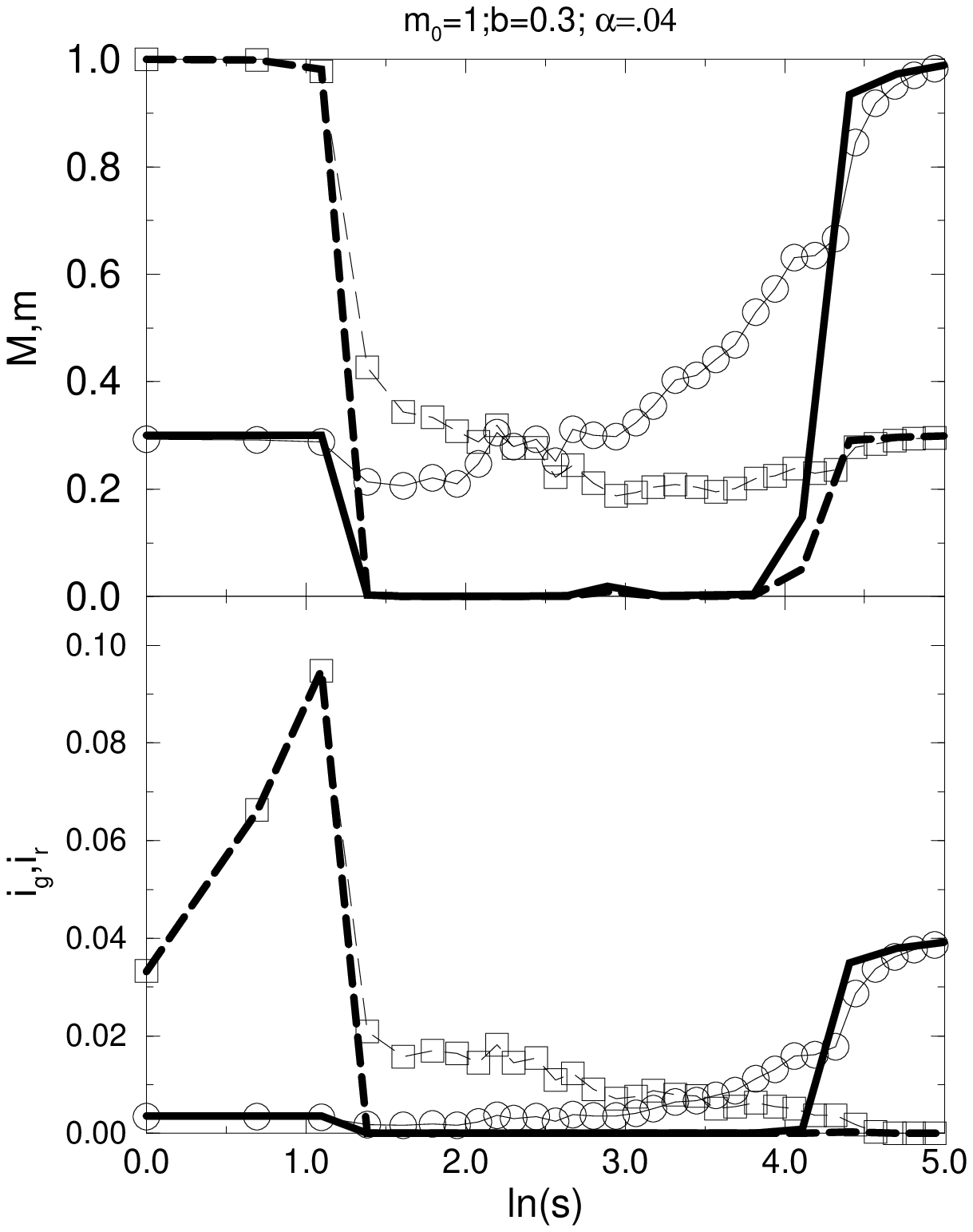}
\end{center}
\caption{ 
Same as Fig.1, for $b=0.3$ and $\alpha=0.04$.
}
\label{2}
\end{figure}

\vspace{-0cm}
\begin{figure}[t]
\begin{center}
\epsfysize=10cm
\leavevmode
\epsfbox[1 1 500 500]{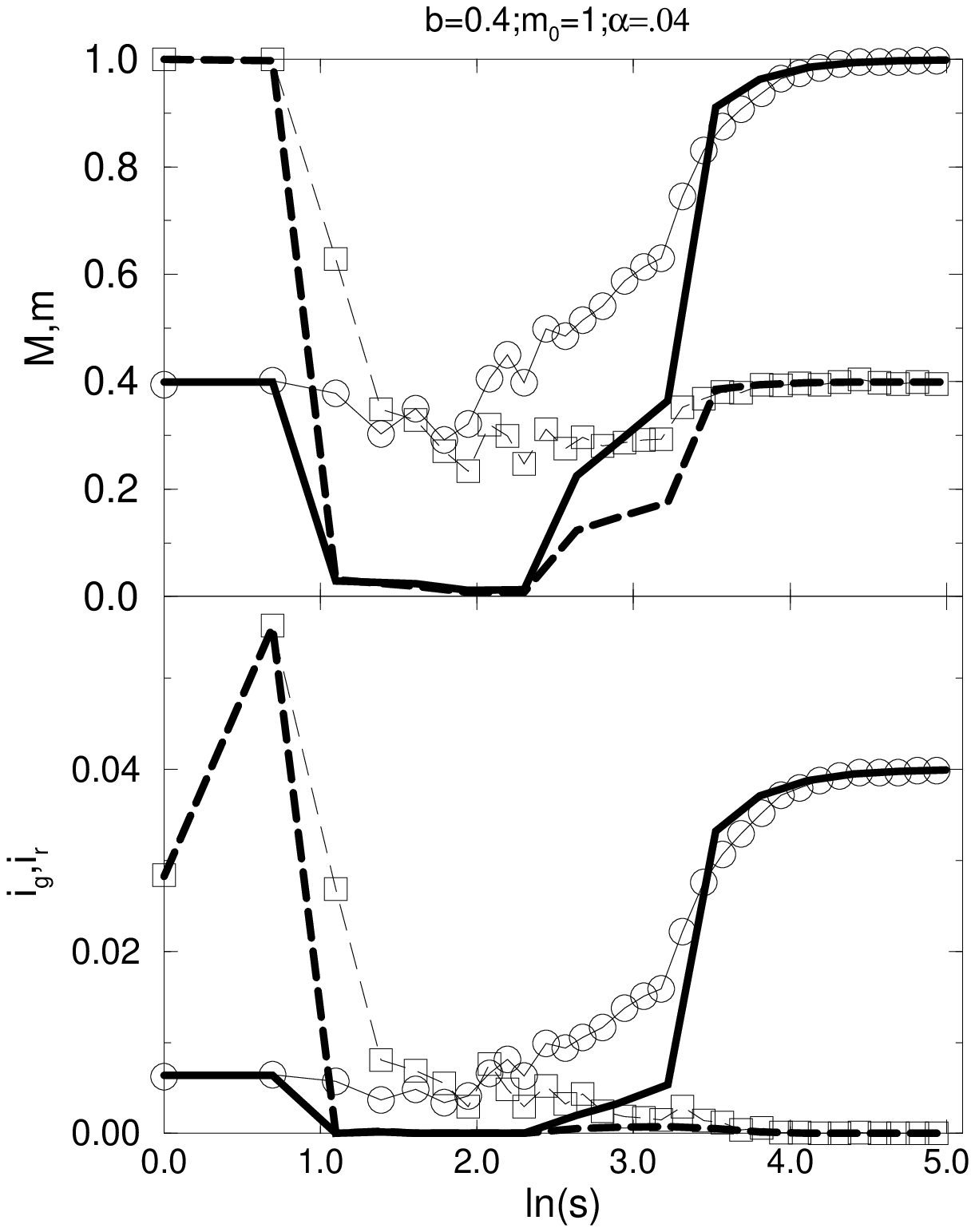}
\end{center}
\caption{  
Same as Fig.1, for $b=0.4$ and $\alpha=0.04$.
}
\label{3}
\end{figure}

\vspace{-0cm}
\begin{figure}[t]
\begin{center}
\epsfysize=9cm
\leavevmode
\epsfbox[1 1 500 500]{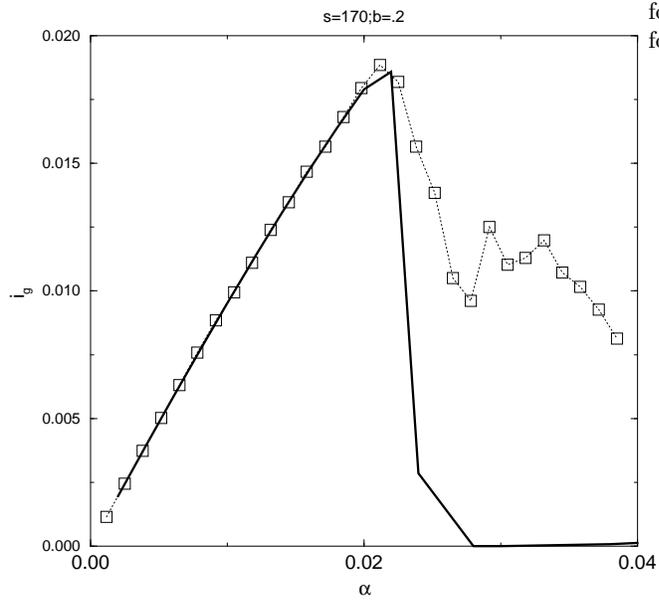}
\end{center}
\caption{ The categorization information as a function of $\alpha$,
for $b=0.2$ and $S=170$.
Asymptotitc (solid) and
Simulation for $N=5,000$ (dashed).      }
\label{4}
\end{figure}

\end{multicols}
\end{document}